\documentclass{article}
\usepackage{amsmath,amssymb}
\newcommand{\ba}{\begin{eqnarray}}
\newcommand{\ea}{\end{eqnarray}}
\usepackage[pdftex]{graphicx}
\usepackage{array,multirow,makecell}
\usepackage{placeins}
\usepackage{setspace}
\usepackage{url,hyperref}
\date{\today}

\makeatletter
\newskip\@bigflushglue \@bigflushglue = -100pt plus 1fil

\def\bigcentering{\let\\\@centercr\rightskip\@bigflushglue%
\leftskip\@bigflushglue
\parindent\z@\parfillskip\z@skip}

\makeatother

\title{\bf Hierarchical PCA and Applications
to Portfolio Management}

\author{Marco Avellaneda\\
Courant Institute of Mathematical Sciences, NYU\\
251 Mercer Street, New York, NY 10012}
\begin{document}

\maketitle

\begin{abstract}
 It is widely known that the common risk-factors derived from PCA beyond the first eigenportfolio are generally difficult to interpret and thus to use in practical portfolio management. We explore a alternative approach (HPCA) which makes strong use of the partition of the market into sectors. We show that this approach leads to no loss of information with respect to PCA in the case of equities (constituents of the S\&P 500) and also that the associated common factors admit simple interpretations. The model can also be used in markets in which the sectors have asynchronous price information, such as single-name credit default swaps, generalizing the works of Cont and Kan (2011) and Ivanov (2016).
\end{abstract}

\section{Introduction}
Principal Components Analysis (PCA) and random matrix theory (RMT) have become widespread tools for data analysis. PCA (Joliffe (2002) \cite{j}) provides a mathematical and objective approach to extract economic information from the correlation matrices of asset returns. In this approach, the analyst extracts common risk factors from the eigenvectors and eigenvalues of the correlation matrix.

The first eigenvector of the correlation of stock returns corresponds to the solution of the variational problem
\begin{equation}
V^{(1)} = argmax\ \Bigl\{ V^t R V ; || V ||=1\Bigr\}.
\label{pca}
\end{equation}
Here, $R$ is the correlation matrix of daily returns and $||.||$ is the Euclidean norm in $ R^n$, $n$ being the total number of assets. Equation \ref{pca} shows that the principal eigenvector is represents the direction (line) which ``captures the most variance''  as described by the correlation matrix.  The first eigenvector satisfies
\begin{equation}
 R V^{(1)} = \lambda^{(1)} V^{(1)}.
\label{eval}
\end{equation}

PCA also finds recursively additional (orthogonal) directions beyond $V^{(1)}$ which capture the most variance. The other eigenvectors and eigenvalues are computed in the same way as Eq. (
\ref{pca})) with the maximization to the sub-space orthogonal to the space spanned by the ones computed previously, i.e,
\begin{equation}
V^{(k)}= argmax\ \Bigl\{ V^t R V ;||V||=1,V^{(k) t}∙V^{(l)}=0,\ 1\leq l<k   \Bigr\}.
\end{equation}
The eigenvalues satisfy $\lambda^{(1)}>\lambda^{(2)}\geq ... \geq \lambda^{(n)}$. Assume that the data corresponds to the daily returns of a group of stocks. The Karhunen-Loeve representation of the standardized returns is

\begin{equation}
X_j = \sum\limits_{k=1}^n \sqrt{\lambda^{(k)}} V_j^{(k)}\, F^{(k)}
\label{kl}
\end{equation}
where
\begin{equation}
F^{(k)} = \frac{1}{\sqrt{\lambda^{(k)}}} \sum\limits^n_{i=1} V^{(k)}_i \, X_i.
\label{factor}
\end{equation}

By construction,  $F^{(k)}$ are uncorrelated and have variance 1. Since these random variables are linear combinations of the daily standardized returns of the assets, we call them (standardized) ``eigenportfolio (EP) returns'', with the caveat that the actual portfolio ``weights'' are obtained by dividing each entry of the eigenvector by the volatility of the asset (Avellaneda and Lee 2008, 2010) \cite{al}.\footnote{ We consider correlations instead of covariances because it mathematically simpler to work in dimensionless units, i.e. to reduce to the case when all the volatilities are equal to one.}

PCA is a framework for learning about the common factors which affect the returns of a given group of assets. The first eigenportfolio, associated with the r.v. $F^{(1)}$, is a common risk factor which explains the maximum variability. We can write a one-factor model for each asset, namely
\begin{equation}
X_j= \beta_j F^{(1)}+ \epsilon_j
\label{reg}
\end{equation}
where $\beta_j$ is the regression coefficient of the standardized return on the first EP. The ``residuals'' $\epsilon_j$ in equation \ref{reg} are uncorrelated with $F^{(1)}$, which is nice. However, they are generally correlated for different stocks.

The regression coefficients satisfy
\begin{equation}
\beta_j = \sqrt{\lambda^{(1)}}\, V^{(1)}_j, \ \  j=1,...,n.
\end{equation}
In the case of economic data, which is noisy, the consensus is to disregard EPs which correspond to low eigenvalues. In a celebrated paper, Laloux {\it et al} (2000) \cite{l} proposed to use random matrix theory (RMT) to establish a cutoff in the number of EPs use to model the standardized returns, namely
\begin{equation}
X_j= \sum\limits^m_{k=1} \beta_j^{(k)}\,  F^{(k)}\ + \epsilon_j
\label{truncation}
\end{equation}
where $\beta_j^{(k)}$ are ``factor loadings’’ and (with a slight abuse of notation) $\epsilon_j$  are residuals obtained after ``defactoring'' relatively to the $m$ eigenportfolios. The number $m$ is a cutoff which is to be determined from the context.

According to \cite{l}, the eigenvalues of a pure noise matrix follow the Marcenko-Pastur distribution and have a spectrum which,
 for large matrices,is  asymptotically bounded from above by $\lambda^{+,MP}=(1+\sqrt{n/T})^2$, where $T$ is the number of observations. Asymptotics should hold in the limit $n/T \rightarrow \gamma$ (a constant) as $n$ and $T$ both tend to infinity. The way to use RMT to calculate the cutoff is to construct the correlation matrix   $R_{i,j}^{(m)}=Corr(\epsilon_i,\epsilon_j )$  for m large enough and verify that its top eigenvalue is of the order of  $\lambda^{+,MP}$. One can also compare the empirical distribution of eigenvalues with the Marcenko-Pastur probability distribution.

PCA aided by RMT is an elegant approach to analyzing correlation matrices of financial data and can also be applied to may areas of science. The main strength of the method is that it can detect common risk factors based on a matrix of asset returns, without any additional information. In other works, PCA ``lets the data speak for itself''.
 Generally speaking, PCA explains the most variability with the smallest number of factors. Most studies tend to justify the PCA approach by recognizing that it produces some factors which have {\it ex-post} economic interpretations, such as equating $EP^{(1)}$ with the Sharpe Market Portfolio (Boyle 2017) \cite{b}, or attempt to interpret higher-order EPs in terms of industry sectors \cite{al}.  In the case of fixed-income, the EPs are often identified with ``parallel shifts'', or with long-term vs short-term oscillations of the yield curve (Litterman and Scheinkman, 1991) \cite{ls}.

\section{The identification problem}
One of the frequent criticisms of PCA in Finance is that the common risk factors generated by higher-order eigenportfolios -- aside from the first eigenportfolio -- are difficult to interpret and appear to be unstable across time. We call this
the {\it identification problem}. Because of it, many portfolio managers favor traditional factor models such as Barra; see Shkolnik {et. al.} (2016) \cite{s} for alternative
approaches to model financial correlations.

The identification problem in PCA reflects the uncertainty, or unreliability, of cross-asset correlations. From a practical point of view, as the size of trading universe increases, the correlations of assets which are not economically related (a tech stock with an energy stock, or with a foreign stock) are difficult to quantify and may be noisy. This could be due to several reasons: the lack of ``explanation'' for the relation between the stocks, or perhaps that their prices are not sampled simultaneously (e.g. if they are end-of-day prices in different time-zones) or that the number of observations is not large compared to the number of assets considered. For example, empirical correlations of price changes of out-of-the money options with different underlying assets may not be as reliable or significant as the data would suggest.

To mitigate the identification problem, we should seek a factor model which can recognize the economic nature or function of the asset as well as the statistical properties of returns. This lead us to the model described hereafter.

\section{Hierarchical PCA}
The hierarchical PCA (HPCA) applies to markets which can be partitioned into several sectors or asset-classes. Consider first an abstract market, in which the empirical data matrix of asset returns, with dimensions $T\times n$, can be partitioned into ``blocks of columns'' labeled $k=1,2,...,b$.  These blocks have dimensions $T \times n_k$ with $k=1,2,...,b$.  Each block represents data sampled from a sector. For simplicity, we assume that the indices of the securities are organized so that blocks which are adjacent to one another in the matrix and do not overlap. We have a few concrete situations in mind:

\begin{itemize}
\item	The blocks represent data of industry sectors for equities in the same economy (e.g. sectors associated with the 500 or so stocks in the S\&P 500 index). In this case, the columns of a block correspond to the historical standardized returns of the stocks in the sector observed over $T$ consecutive dates.

\item	Each block represents a stock or index and all of the derivatives written on it. In this case, the columns in a block represent the returns of the stock and the changes of the implied volatilities of options with different strikes and tenors written on the stock (Dobi 2015 \cite{d}).

\item In the context of credit derivatives, the data represents changes in credit spreads for CDS. The blocks correspond to CDS referencing the same obligor (issuer) but with different tenors (Cont and Kan (2011) \cite{ck}, Ivanov (2017) \cite{i}).
\end{itemize}

Define the function $I(j)=k$ if asset $j$ is in block $k$. According to Eq. (\ref{kl}) we can write, for each asset in the ``big universe'',
\begin{equation}
X_j= \beta_j \, F^{(1, I(j))}+  \epsilon_j,
\label{1factor}
\end{equation}
where $\beta_j$ is the regression coefficient of the returns of asset $j$ on the first factor of block $I(j)$ and $\epsilon_j$ is the residual.

We shall make the following assumption (``HPCA assumption''):
\begin{equation}
\boxed{ \hbox{If}\ I(i)\neq I(j),\ \hbox{then}\     Corr(\epsilon_i,\epsilon_j)=0.}
\label{hpca}
\end{equation}
The assumption states that residuals are uncorrelated if their assets belong to different sectors. Equation (\ref{1factor}) defines the asset statistics within each block exactly,  and the model is completed by specifying the joint statistics of the factors  $F^{(1,k)},\ k=1,2,…,b.$ The HPCA assumption says nothing new regarding intra-block correlations, which are set equal to the empirical correlations between asset returns within the same sector or block. Of course, the intra-block correlations could be further denoised using RMT if necessary ( \cite{d}).

Using the HPCA assumption Eq. (\ref{hpca}), the proposed model has the modified correlation matrix for asset returns:

\begin{eqnarray}
\nonumber \tilde{R}_{ij} & = & R_{ij}  \  \hbox{if}\  \ I(i) = I(j)\\
 & = & \beta_i \,\beta_j\, \overline{\rho}^{I(i)  I(j)}  \ \hbox{if}\  \  I(i) \neq I(j)
\label{hpcacorr}
\end{eqnarray}
where $\overline{\rho}^{k, k'} = Corr(F^{(1, k)},F^{(1, k')} )$.

\noindent {\bf Proposition 1} {\it Eq. (\ref{hpcacorr}) corresponds to a symmetric non-negative matrix with $\tilde{R}_{ii} =1$ for all $i$. In particular, it corresponds to the correlation matrix of a system of standardized random variables.}

\noindent {\bf Proof}. To check non-negative definiteness, note that for all $\theta \ \in \ R^n$ we have
\begin{equation}
\theta^t\ \tilde{R} \theta = \sum\limits^b_{k=1} \sum\limits_{I(i)=I(j)=k} \theta_i\theta_j ( R_{i j} - \beta_i \beta_j ) +
\sum\limits_{k, k'=1}^b (\sum_{I(i)=k}\theta_i \beta_i) \, (\sum_{I(j)=k'}\theta_j \beta_j) \  \overline{\rho}^{k,k'}.
\end{equation}
For any $k$, the matrix $R_{i j} - \beta_i \beta_j$ restricted to sector $k$ is identical to the sector correlation, except for the fact that the
eigenvalue corresponding to $V^{(1,k)}$ is set to zero. In particular, it is non-negative definite. Moreover, the matrix $\overline{\rho}^{k,k'}$ is also a correlation matrix, so it is non-negative definite. Since both summands are non-negative it follows that $ \theta^t\ \tilde{R} \theta \geq 0$ for all $\theta \, \in \, R^n$.

 Since $\tilde{R}$ is non-negative definite, the HPCA assumption is compatible with a multivariate distribution (data model), which presents an alternative to the classical PCA ( Eq. (\ref{truncation})). Tt has a tree structure: in the equity example discussed below, the top vertex corresponds to the ``market''; there are 11 branches corresponding to industry sectors, and each of the 11 vertices has branches corresponding to the stocks in each sector.

 Hierarchical models with more than two layers arise naturally. For instance, HPCA can be used to model ``world portfolios'', in which the first layer consists of countries or regions, the second to industry sector indices in each country. A third ``layer'' could describe the securities in each region/sector.
 
Consider the case of a stock market in which stocks belong to different industry sectors, and then, along with stock returns, include columns associated with equity options returns. In this case, the tree has three layers because we can associate to each stock an additional sub-group: the
block consisting of the returns of implied volatilities (on a constant delta/time-to-maturity grid) and the stock returns. Now the root corresponds to the full market, the first layer corresponds to industry sectors, the second layer corresponds to a stock viewed as an underlying asset and the third layer represents an individual name with all the associated option-implied volatilites.

A similar approach works for credit derivatives. In this case, the returns of the CDS with different tenors referencing each obligor constitute a block associated with an obligor. These blocks can be grouped by industry sectors or, alternatively, blocks could be generated according to membership in a credit index (CCX.IG, CDX.HY, CDX.HV), or both.

In summary, if financial data can be grouped into blocks or sectors with clear economic interpretation, with multiple instruments associated with each block, we can generate a data model with tree-like structure from the HPCA assumption in Eq. (\ref{hpca}). This approach combines information available for each asset (sector, sub-sector, reference obligor, option underlying asset) with the explanatory power of PCA. For simplicity, we will consider the analysis of a two-layer HPCA. Adding more layers is mathematically straightforward.

\section{Spectral analysis}
The HPCA assumption Eq. (\ref{hpca})gives rise to explicitly computable eigenvalues and eigenvectors for the
matrix $\tilde{R}$ defined in Eq. (\ref{hpcacorr}).

\noindent {\bf Proposition 2.} {\it
\begin{enumerate}
\item For each sector $k=1,...,b$, let $\lambda^{(1,k)}\ > \ \lambda^{(2,k)} \geq \ ...\ \geq \lambda^{(n_k,k)}$ denote the $n_k$ eigenvectors of the sector correlation matrix, ordered from largest to smallest, and let $V^{(i,k)}$ be the corresponding eigenvectors. Define the n-dimensional vectors
\begin{eqnarray}
\nonumber W^{(i,k)}_j & = & V^{(i,k)}_j \ \ \hbox{if}\ \ I(j) = k\\
 & = & 0 \ \ \ \ \hbox{if}\ \ I(j) \neq k,
 \end{eqnarray}
which correspond to the embedding of the sector-level eigenvectors, $V^{(i,k)}\ \in \ R^{n_k}$, into the large space $R^n$. The vectors $W^{(i,k)}, \  i=1,..., n_k, \ \ k=1,...,b$ form an orthogonal basis of $R^n$.
\item The subspace $\Omega$ of $R^n$ generated by the vectors $W^{(1,k)},\ k=1,...,b$, {\it viz.}
\begin{equation}
\Omega= \Bigl\{ \sum\limits^b_{k=1} \alpha_k W^{(1,k)} :  (\alpha_1,...,\alpha_b) \in R^b \Bigr\},
\label{omega}
\end{equation}
is invariant under the action of $\tilde{R}$ viewed as an operator from $R^n$ to $R^n$.
\item Consider the $b \times b$ matrix
\begin{equation}
M^{k,k'} := \sqrt{\lambda^{(1,k)}} \sqrt{\lambda^{(1,k')}} \overline{\rho}^{k, k'}.
\label{alpha}
\end{equation}
Let $\mu^{(1)},...,\mu^{(b)}$ denote the eigenvalues of $M$, ranked in decreasing order, and
let $(\alpha^{(k)}=(\alpha^{(k)}_1,....,\alpha^{k)}_b)\ k=1,...,b$ represent the corresponding normalized eigenvectors (defined up to sign).
The vectors
\begin{equation}
\tilde{W}^{(1,k)}=\sum\limits_{p=1}^b \alpha^{(k)}_p\,W^{(l,p)}\
\end{equation}
are eigenvectors of $\tilde{R}$, with corresponding eigenvalues $\mu^{(k)}$, for $k=1,...,b$.
\item For each sector $k$ and each $j,\ 2 \leq j\  \leq n_k$, the vector $W^{(j,k)}$ is an eigenvector of $\tilde{R}$, with eigenvalue $\lambda^{(j,k)}$.
\end{enumerate}
}
This proposition completely characterizes the eigenvalues and eigenvectors of the HPCA correlation matrix relating them to the eigenvalues and eigenvectors of sector PCAs.\footnote{The proof of Proposition 2 is straightforward: one just has to observe that $\beta_i=\sqrt{\alpha^{I(i)}} V^{(1,I(i))}_i$ and calculate explicitly the action of $\tilde{R}$ on each of the vectors $W^{(j,k)}$.} Thus, the HPCA assumption eliminates the identification problem for common factors: ``eigenportfolios'' have concrete meanings attached
 to the information about the correlations of sectors. In the examples to follow, we shall compare HPCA with PCA and show that the former
 is an excellent substitute for the full empirical correlation matrices when we model multivariate financial data.

\section{Application: S\&P 500 constituents}
 We consider  data for $n=434$ equities which are constituents of the S\&P500 index. The data ranges from February 22, 2012 to February 16, 2018. We consider the correlation matrix of standardized stock returns, and define the sectors as General Industry Classification groups (GICs), so $b=11$; see Table 1.

\begin{table}[htbp]
\centering
\title{\bf General Industry Classification (GIC) Sectors}\\
\vspace{0.3cm}
\begin{tabular}{| c | c | c |}
\hline
\vspace{0.1cm}
GIC ($k$)	&	 Description	&	Number of companies($n_k$) \\
\hline\hline
1	& Consumer Discretionary	& 73\\
2	&  Consumer Staples & 	56\\
3	& Energy & 	27 \\
4	& Financials & 	59\\
5	& Health Care & 	51\\
6	& Industrials & 	57\\
7	& Information Technology & 	58\\
8	& Materials & 	23\\
9	& Real Estate	 & 27\\
10	& Telecommunication Services & 	3\\
11	& Utilities & 	28\\
\hline
\end{tabular}
\caption{GIC sectors and number of companies in each sector.}
\label{table:gic}
\end{table}

\subsection{Eigenvalues}
\begin{figure}[htbp]
\begin{bigcenter}
\includegraphics[scale=0.30]{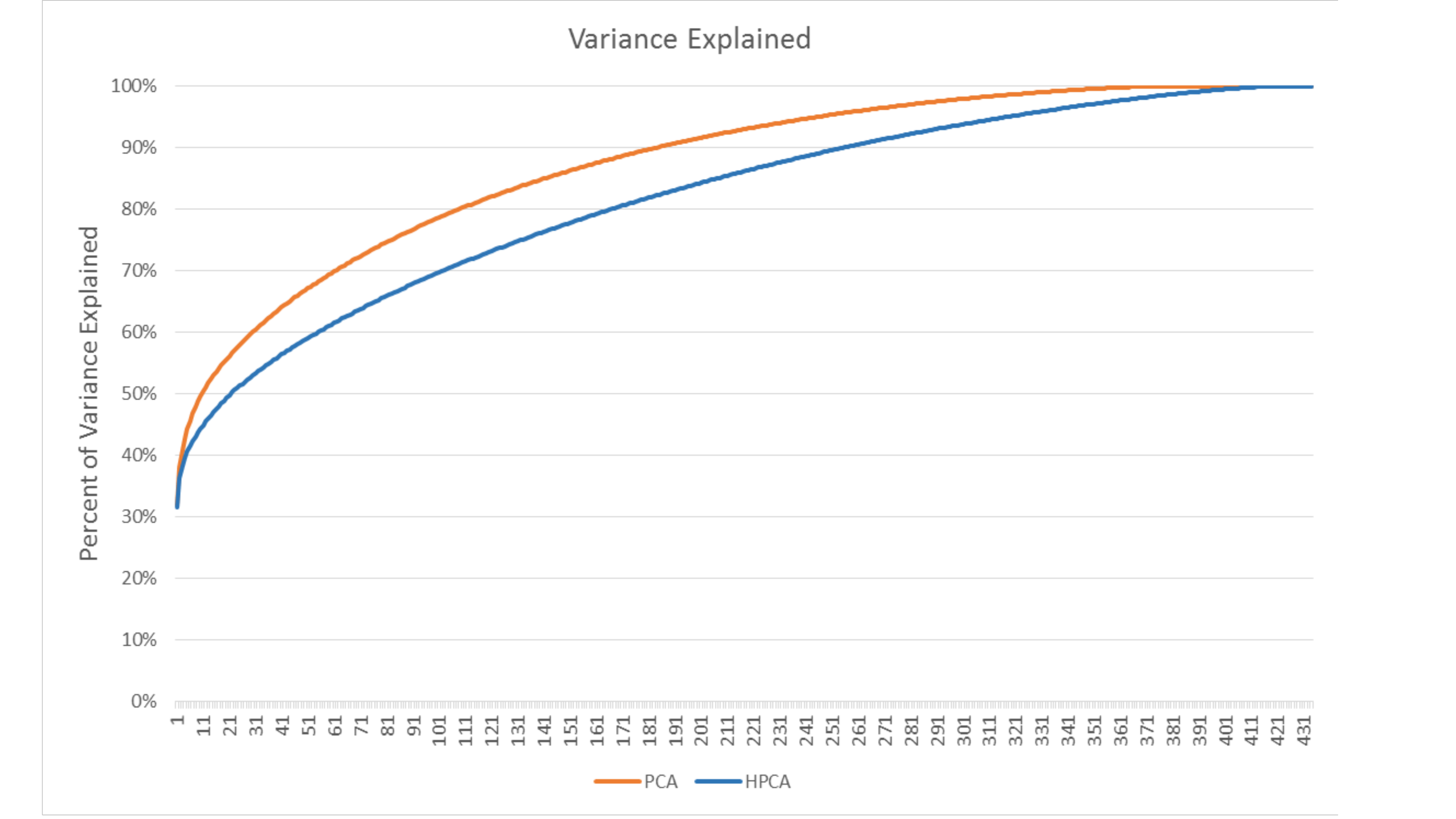} %, width=15 cm
\caption{X=axis: rank ($k$) of the eigenvalues, sorted in decreasing order. Y-axis: sum of the first $k$ eigenvalues divided by $n=434$. The PCA curve rises faster than HPCA, due to the nature of the PCA algorithm.}
\label{variance}
\end{bigcenter}
\end{figure}

We considered the full empirical correlation matrix\footnote{In the sequel we refer to the full empirical correlation matrix as the ``PCA matrix'', for short.} and the HPCA correlation matrix $\tilde{R}$ (``HPCA matrix''). The spectrum of the HPCA matrix is very similar than the one of the empirical correlation matrix $R$, with the difference that the latter eigenvalues at the top of the spectrum are slightly larger the eigenvalues of the HPCA matrix. This is due to the fact that PCA explains more variance with fewer
common factors (see Figure (\ref{variance})). On the other hand, the sum of eigenvalues is equal to $n=434$ in both cases, which means that for high enough rank, the higher-order eigenvalues of HPCA are larger than those of PCA. The lowest eigenvalues of $R$ are infinitesimal, and the latter matrix is degenerate. At the bottom of the spectrum (not shown here) the HPCA spectrum has much higher eigenvalues (separated
from zero) than PCA, since they are bounded from below by the lowest eigenvalue from all the sectors. Thus, the HPCA matrix is better conditioned than
the full empirical matrix.

\begin{table}[htbp]
\centering
\title{\bf Top 25 Eigenvalues of PCA, HPCA and Interpretation}\\
\vspace{0.3cm}
\begin{tabular}{| c | c | c | c| c | c | c |}
\hline
\vspace{0.1cm}
PCA	&	HPCA	&	Eigenportfolio	&	&	PCA	&	HPCA	&	Eigenportfolio	\\
\hline
\hline
138.87	&	137.19	&	{\bf Multi-sector}	&	&	2.79	&	2.18	&	Industrials	\\
26.84	&	20.70	&	{\bf Multi-sector}	&	&	2.52	&	2.15	&	Consumer Disc.	\\
11.88	&	8.18	&	{\bf Multi-sector}	&	&	2.46	&	2.14	&	Healthcare	\\
7.70	&	5.91	&	{\bf Multi-sector}	&	&	2.36	&	2.09	&	Inf. Technology	\\
6.87	&	4.93	&	{\bf Multi-sector}	&	&	2.32	&	2.03	&	{\bf Multi-sector}	\\
5.75	&	3.69	&	{\bf Multi-sector}	&	&	2.24	&	1.94	&	Technology	\\
5.16	&	3.38	&	Consumer Disc.	&	&	2.20	&	1.93	&	Industrials	\\
4.70	&	2.88	&	{\bf Multi-sector}	&	&	2.18	&	1.92	&	Energy	\\
3.90	&	2.80	&	Financials	&	&	2.13	&	1.80	&	Consumer Disc.	\\
3.61	&	2.68	&	{\bf Multi-sector}	&	&	2.06	&	1.59	&	Inf. Technology	\\
3.48	&	2.67	&	Healthcare	&	&	2.01	&	1.57	&	Industrials	\\
3.02	&	2.53	&	Consumer Disc.	&	&	1.96	&	1.57	&	Healthcare	\\
2.87	&	2.25	&	Healthcare	&	&		&		&		\\
\hline
\end{tabular}
\caption{Top 25 eigenvalues of PCA and HPCA, sorted in decreasing order. The column ``Eigenportfolio'' gives an interpretation of the corresponding HPCA
eigenportfolio. ``Multi-sector'' corresponds to a $\mu^{(k)}$-eigenvalue and eigenvector, which are combinations of the {\it first} eigenportfolios for each of the 11 sectors (space $\Omega$). The other eigenvalues/eigenvectors correspond to higher-order eigenvalues/eigenvectors for individual GIC sectors. Notice that, after sorting, some of the GIC eigenportfolios are more important in terms of explaining variability than the higher-order multi-sector portfolios.}
\label{table:eigenvalues}
\end{table}
\newpage
\subsection{Eigenvectors}
\begin{figure}[htbp]
\begin{bigcenter}
\includegraphics[scale=0.30]{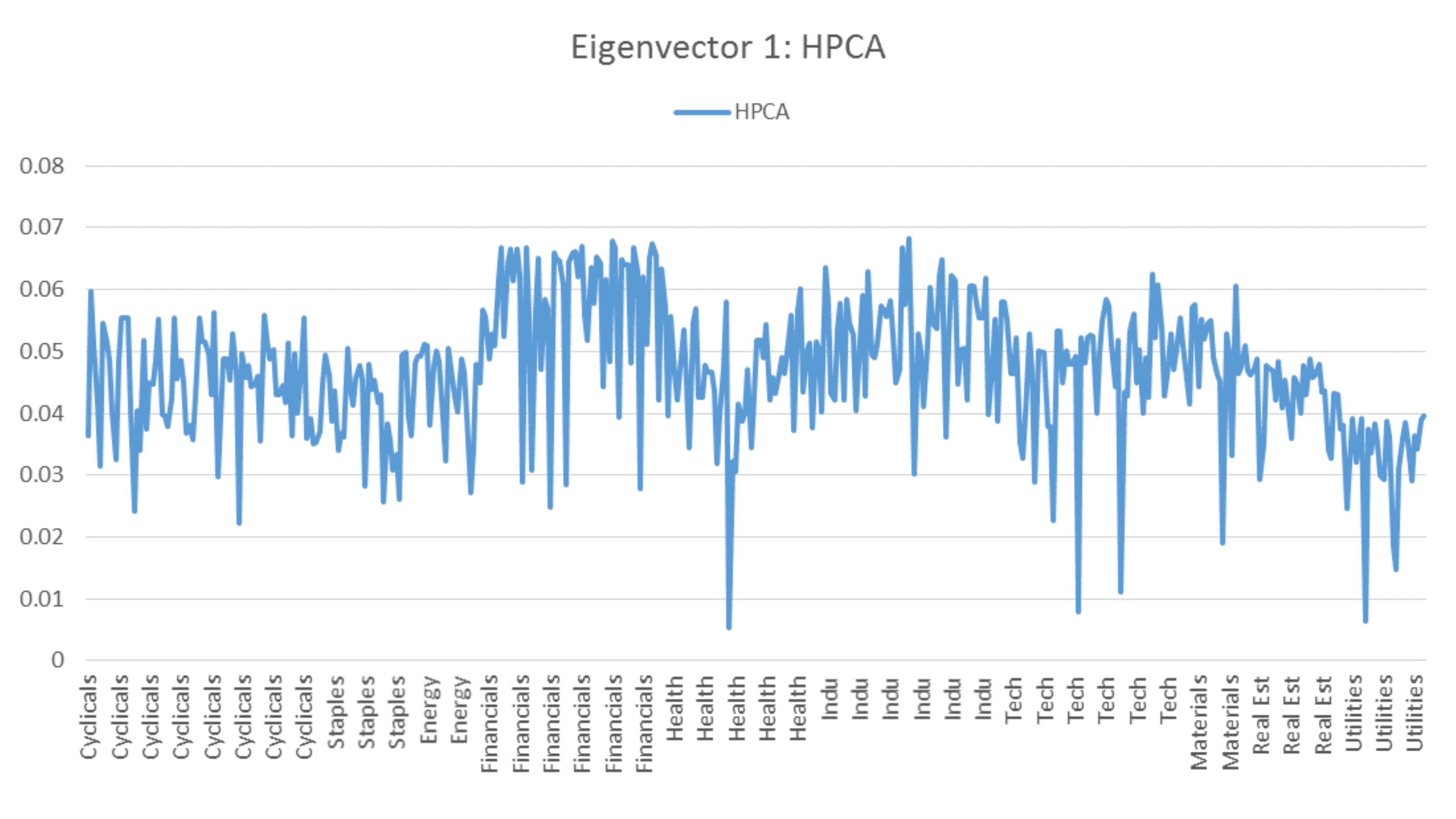} %, width=15 cm
\caption{First eigenvector of HPCA. Variance explained= 30\%.}
\label{hpca1}
\end{bigcenter}
\end{figure}

\begin{figure}[htbp]
\begin{bigcenter}
\includegraphics[scale=0.30]{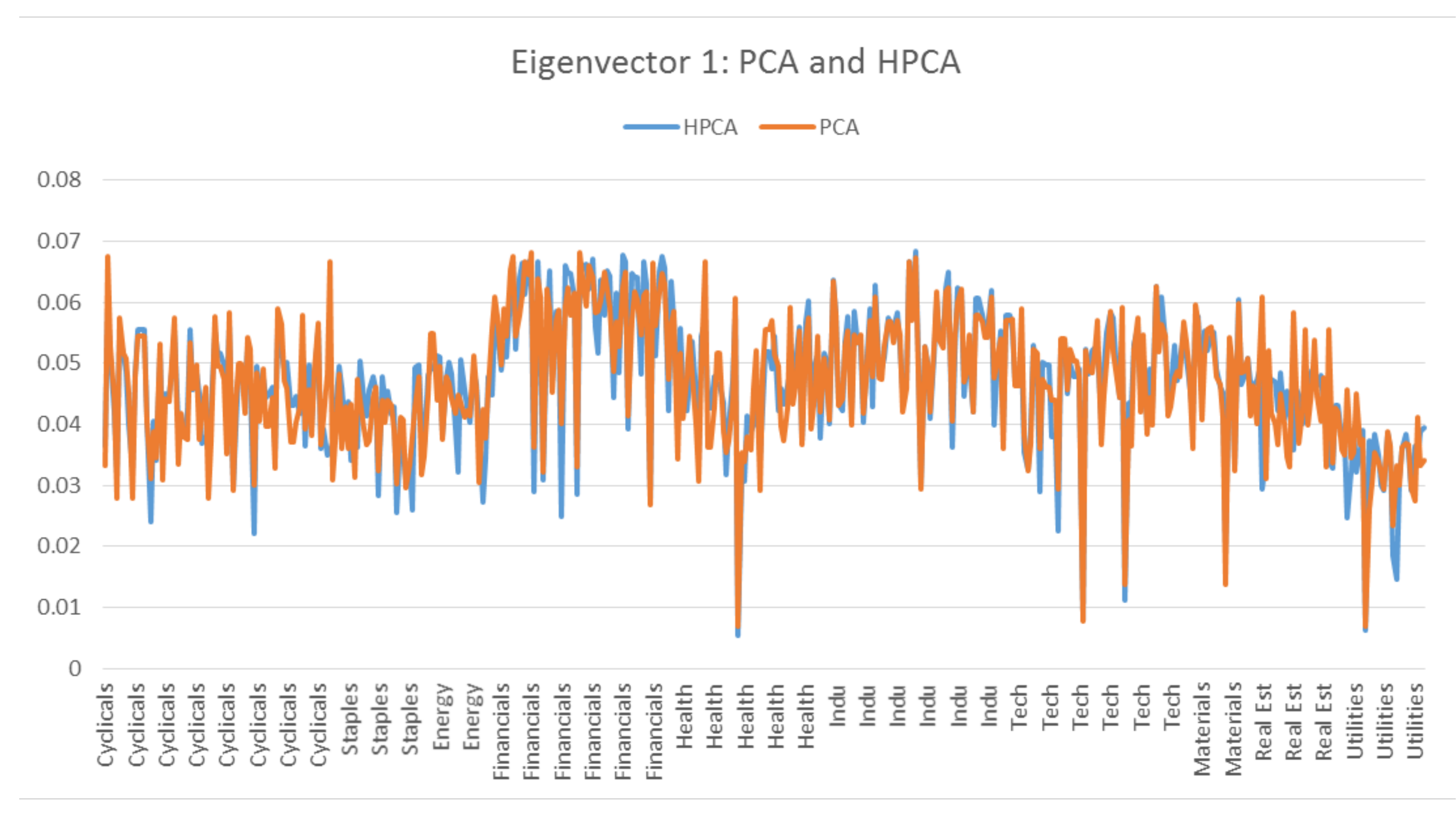} %, width=15 cm
\caption{Comparison of the first eigenvectors of HPCA and PCA, which have approximately the same explanatory value. Their Euclidean distance (RMS error) is
$5.5 \times 10^{-3}$, which is an order of magnitude smaller than the average entry size.}
\label{ev1}
\end{bigcenter}
\end{figure}

We turn to empirical analysis of the eigenvectors of the HPCA and the empirical correlation matrices, {\it i.e.} to the issue of identification problem
for PCA/HPCA. The first eigenvectors for HPCA and PCA are plotted in Figures (\ref{hpca1}) and (\ref{ev1}). Since the first eigenvector of $M$ has positive entries and the first eigenvectors of sector correlations also have positive entries due to the positive correlations of stocks ( \cite{al},\cite{b} ; EV1 loadings are positive for both PCA and PCA. Figure (\ref{ev1}) superimposes both eigenvectors. The ordering of the X-axis is alphabetical in each sector and sectors are grouped displayed in increasing order of GIC according to Table (\ref{table:gic}). The two eigenvectors are practically indistinguishable in the sense that their average difference is of order $1.0\times 10^{-5}$ and the standard deviation (centered RMS distance) is $5.3 \times 10^{-3}$. The RMS error is one order of magnitude smaller than the average size of each entry in the eigenvectors which is approximately equal to $4.7 \times 10^{-2}$, in both cases.

This identifies the first eigenportfolio of the market as a ``portfolio of first eigenportfolios''  of different sectors (GICs). The difference in explanatory power between the two eigenvectors is the difference between the corresponding eigenvalues, divided by the number of stocks, namely $(138.87-137.19)/434=0.39\%$, which is negligible in this context. In particular, this suggests that using the first HPCA eigenportfolio as a proxy for the market portfolio gives rise to a better description of the market portfolio and an easier way to allocate to each stock. For instance, the first EV could be proxied by a capitalization-weighted sector ETF.\footnote{A careful analysis of this idea, including out-of-sample tracking error analysis, will be done in a separate publication.}.

\begin{figure}[htbp]
\begin{bigcenter}
\includegraphics[scale=0.30]{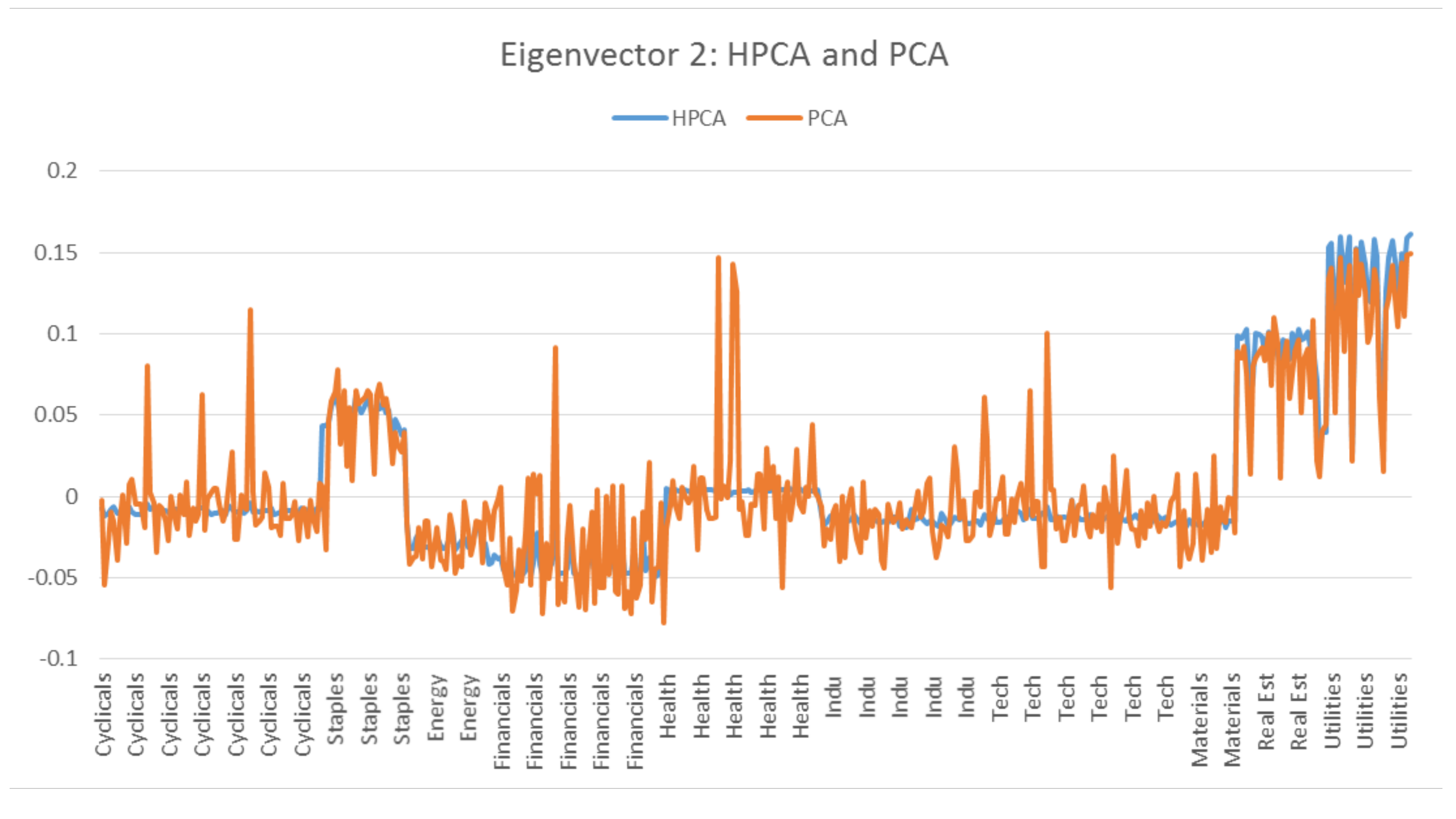} %, width=15 cm
\caption{Comparison of the second eigenvectors. The PCA eigenvector is essentially a noisy version of the HPCA eigenvector.}
\label{ev2}
\end{bigcenter}
\end{figure}

\begin{figure}[htbp]
\begin{bigcenter}
\includegraphics[scale=0.60]{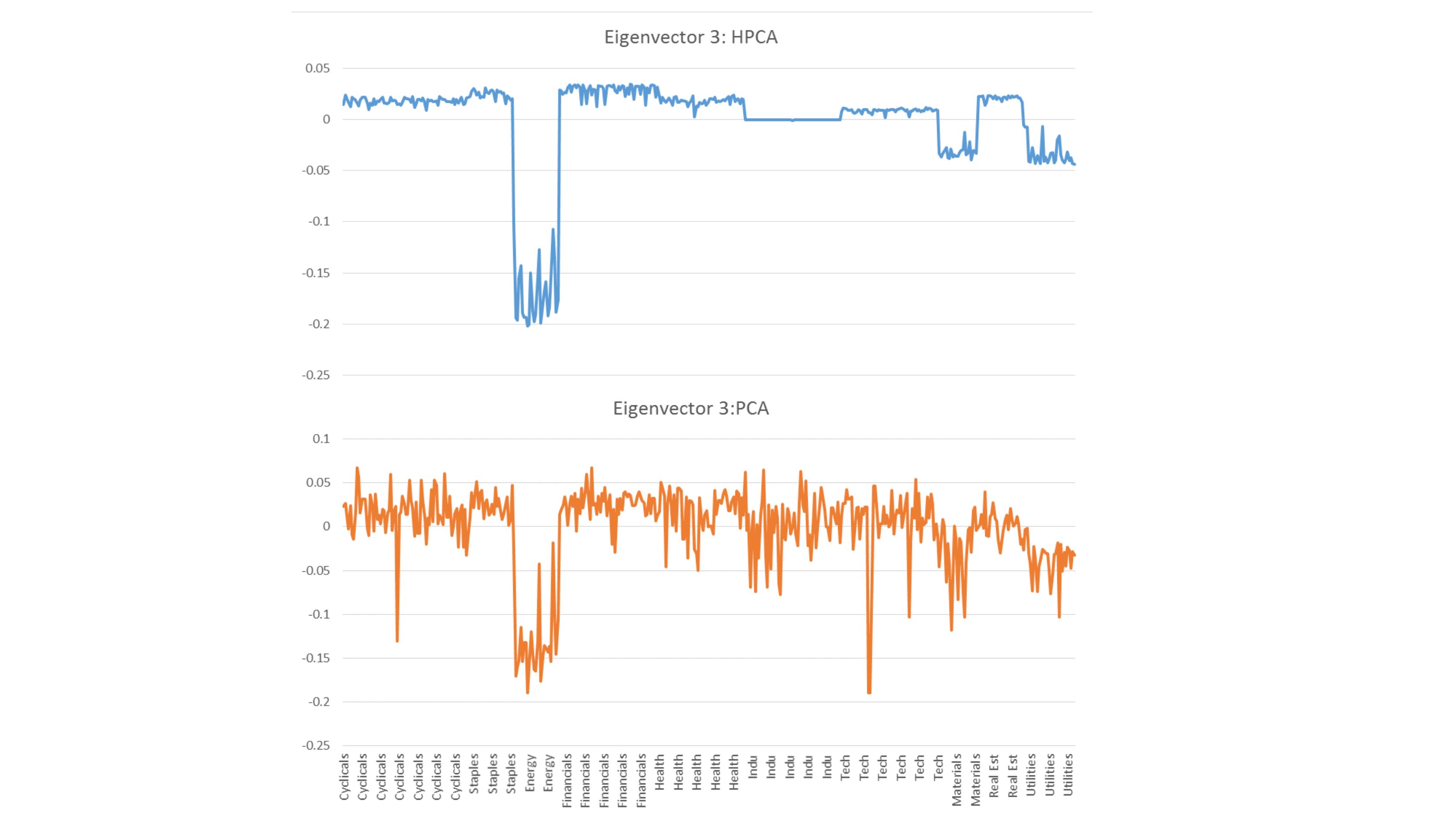} %, width=15 cm
\caption{The third eigenvectors of HPCA: one can observe again that PCA EV3 is a noisy version of HPCA EV3.}
\label{ev3}
\end{bigcenter}
\end{figure}

\begin{figure}[htbp]
\begin{bigcenter}
\includegraphics[scale=0.60]{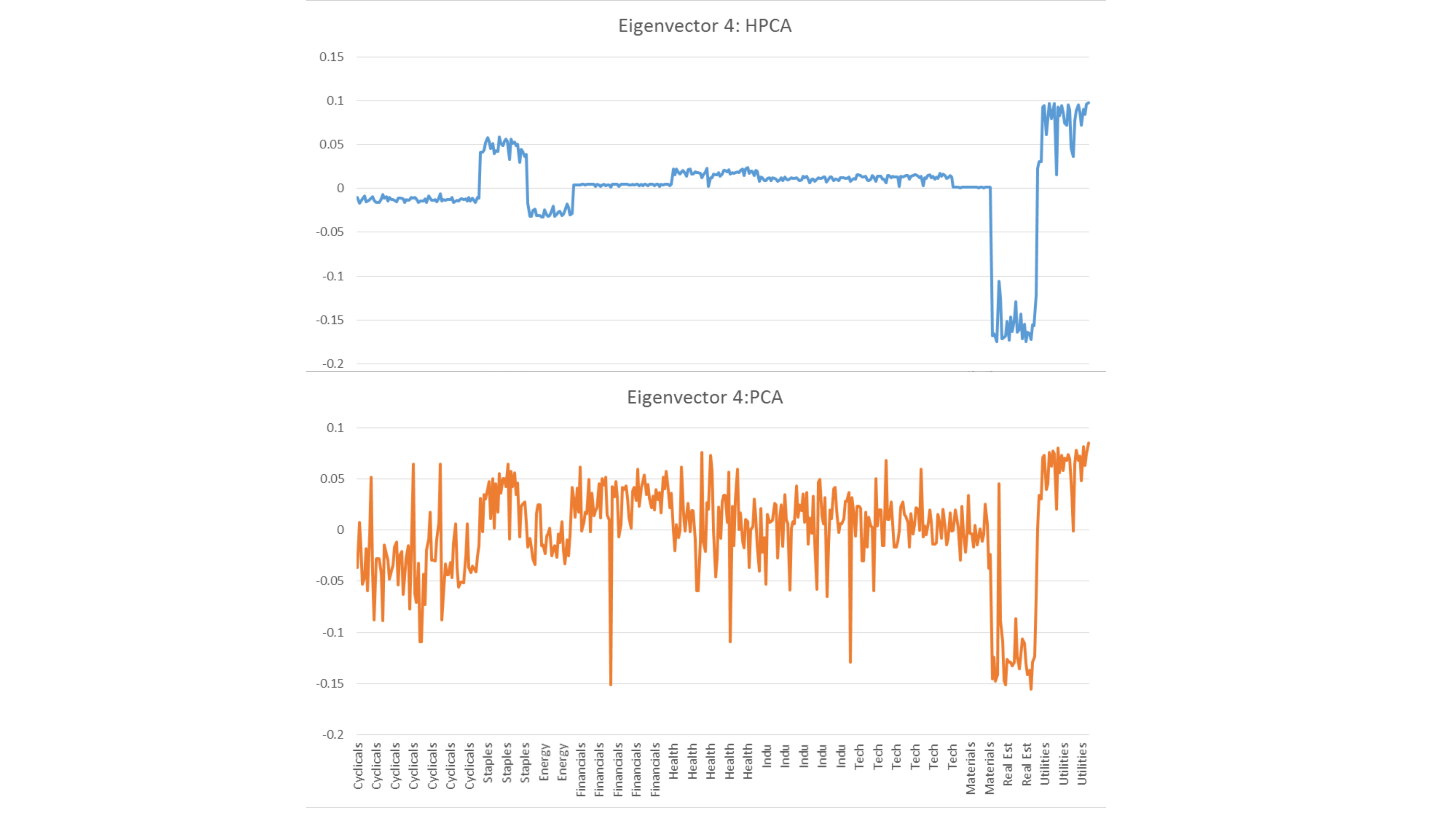} %, width=15 cm
\caption{The fourth eigenvectors. Notice the similar loadings for sectors.}
\label{ev4}
\end{bigcenter}
\end{figure}

\begin{figure}[htbp]
\begin{bigcenter}
\includegraphics[scale=0.60]{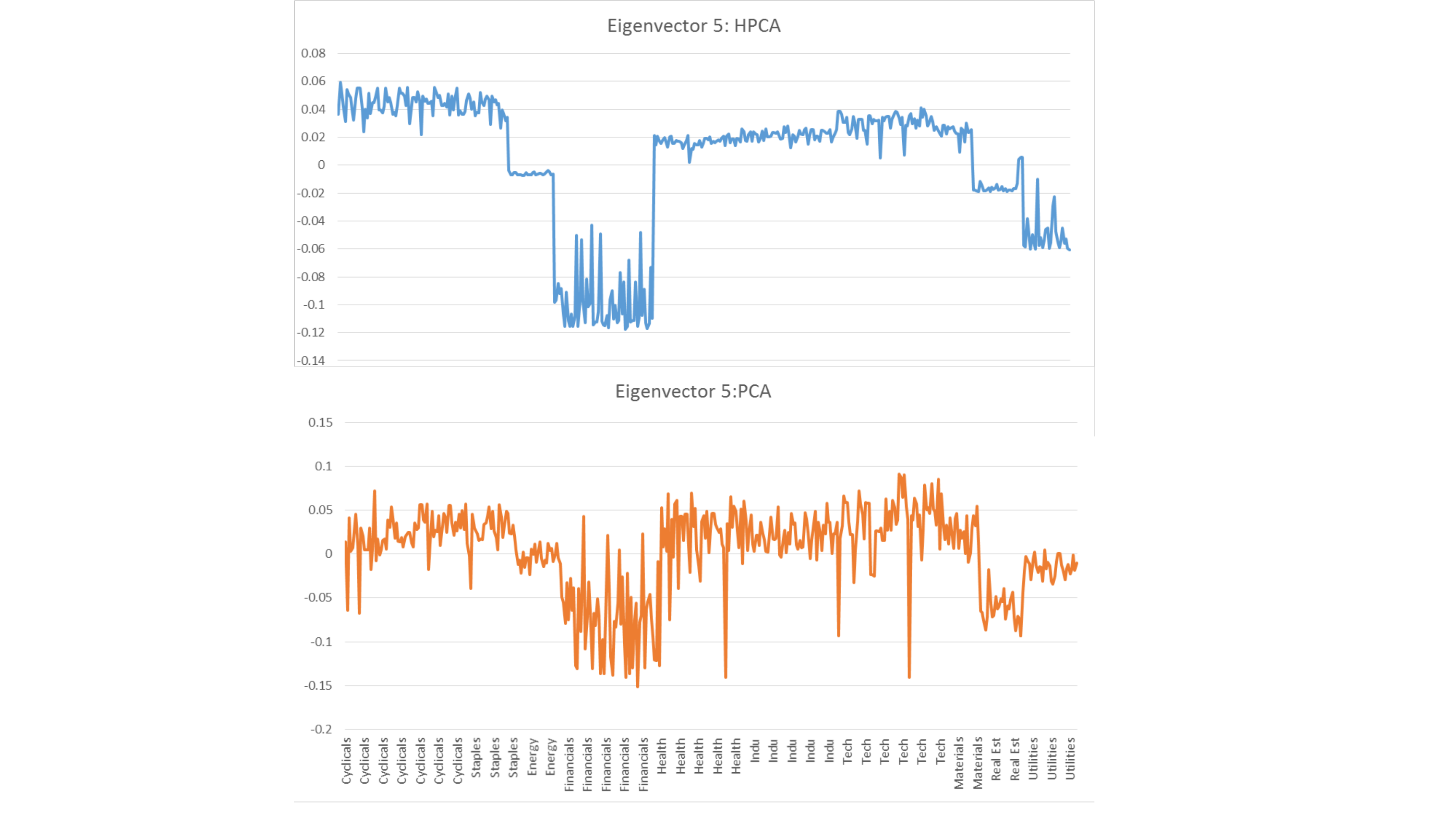} %, width=15 cm
\caption{The fifth eigenvectors. Notice the similar loadings for sectors.}
\label{ev5}
\end{bigcenter}
\end{figure}

\begin{figure}[htbp]
\begin{bigcenter}
\includegraphics[scale=0.60]{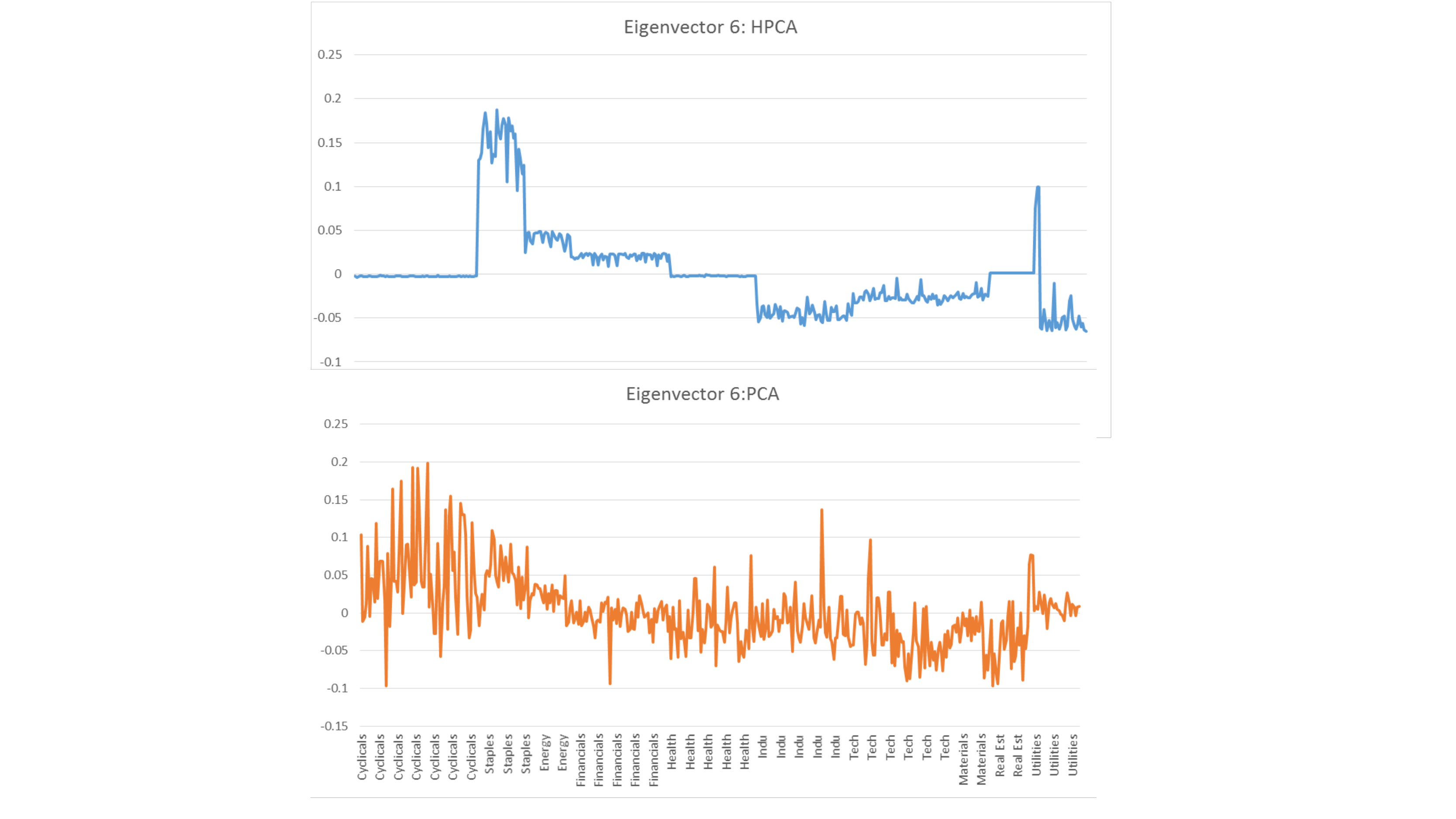} %, width=15 cm
\caption{The sixth eigenvectors. In this case, PCA presents a different shape and is not ``localized'' on any sector. The leftmost
part of the PCA eigenvector corresponds to Consumer Discretionary.}
\label{ev6}
\end{bigcenter}
\end{figure}

\begin{figure}[htbp]
\begin{bigcenter}
\includegraphics[scale=0.60]{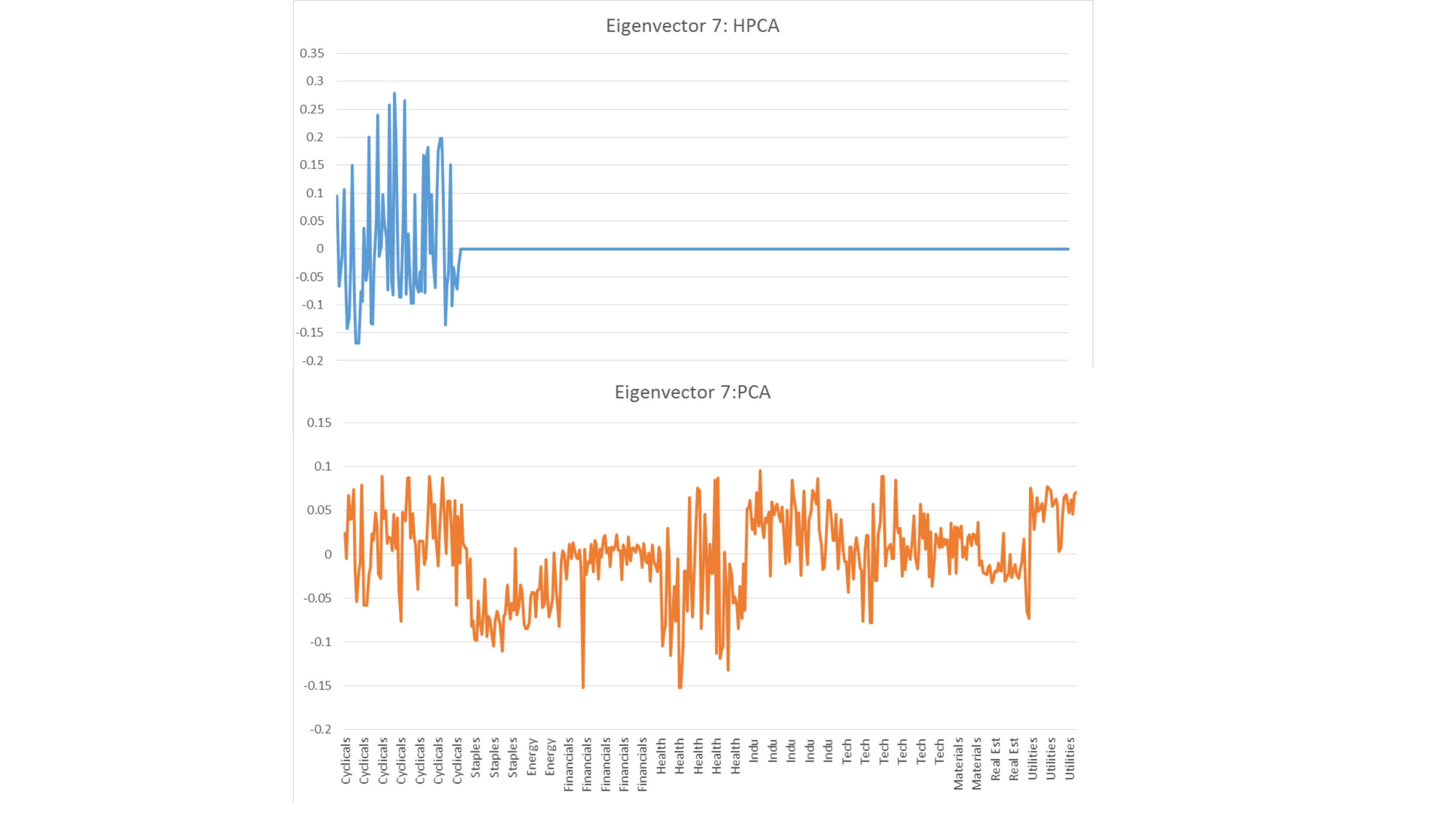} %, width=15 cm
\caption{The seventh eigenvectors. The HPCA is essentially an eigenvector localized on the Consumer Discretionary sector (the second eigenvector). The PCA eigenvector is completely delocalized.}
\label{ev7}
\end{bigcenter}
\end{figure}

\begin{figure}[htbp]
\begin{bigcenter}
\includegraphics[scale=0.60]{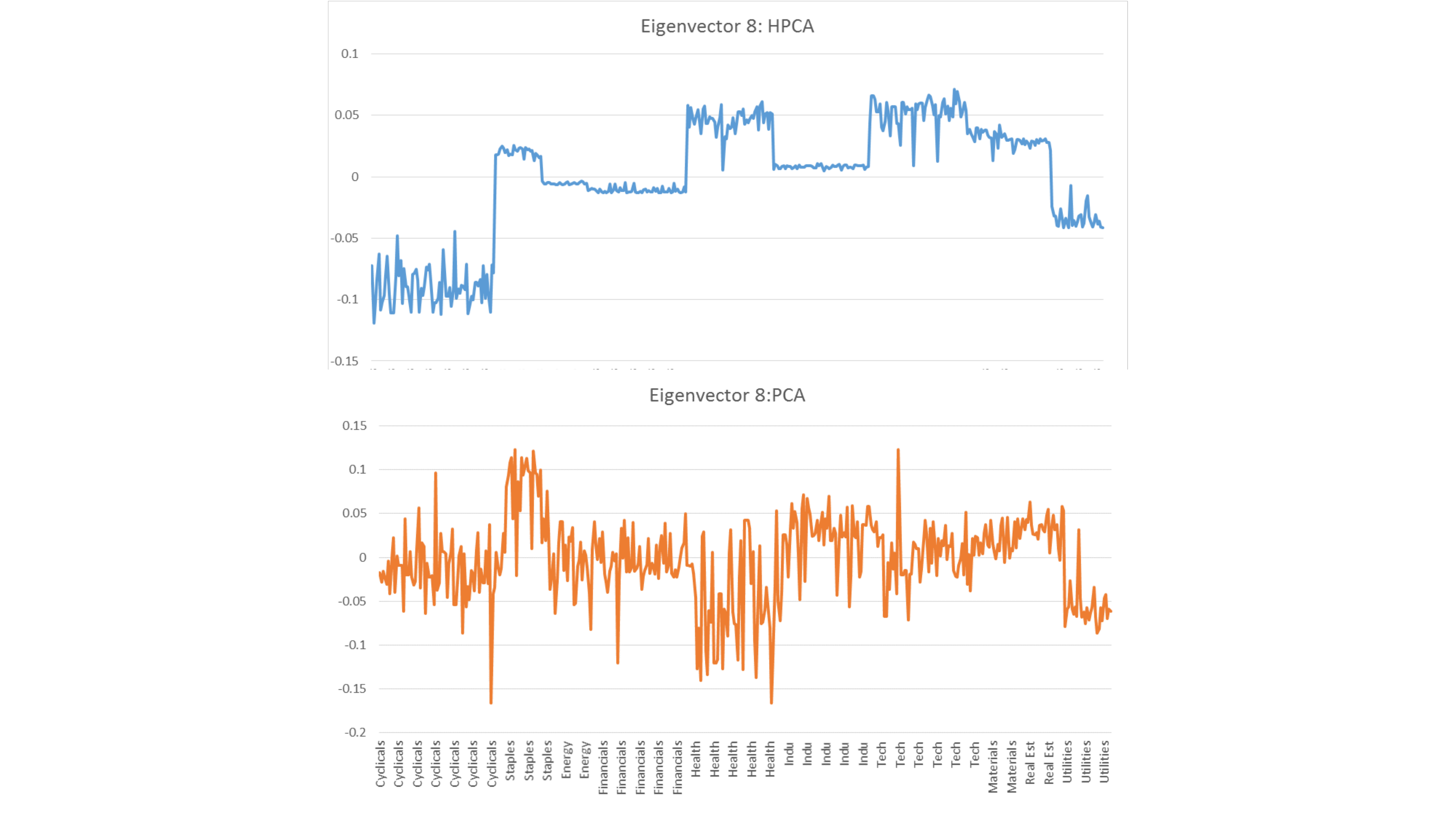} %, width=15 cm
\caption{Eight eigenvectors.}
\label{ev8}
\end{bigcenter}
\end{figure}

\begin{figure}[htbp]
\begin{bigcenter}
\includegraphics[scale=0.60]{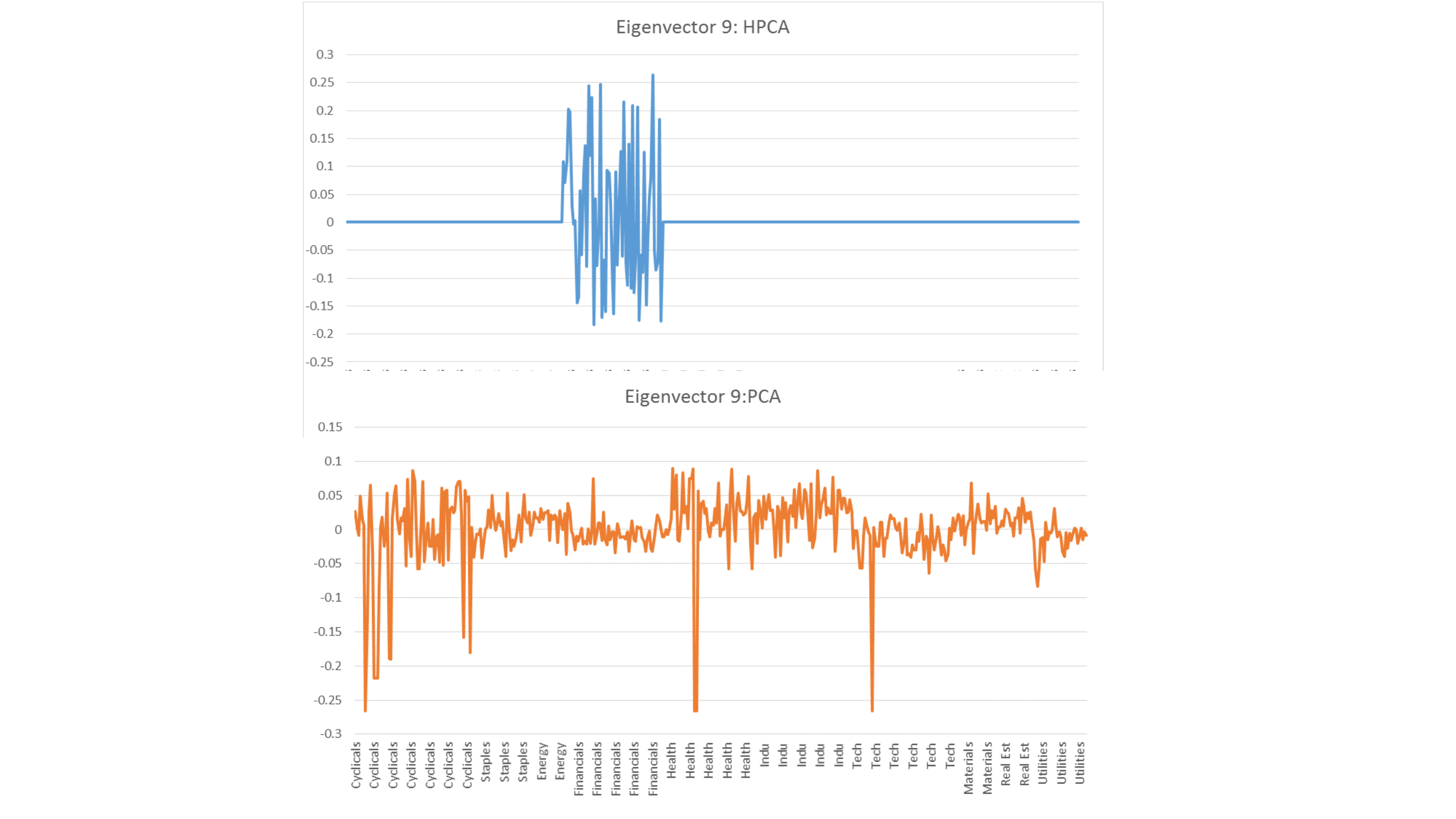} %, width=15 cm
\caption{Ninth eigenvectors. The HPCA eigenvector is localized in the Financials sector.}
\label{ev9}
\end{bigcenter}
\end{figure}

\begin{figure}[htbp]
\begin{bigcenter}
\includegraphics[scale=0.60]{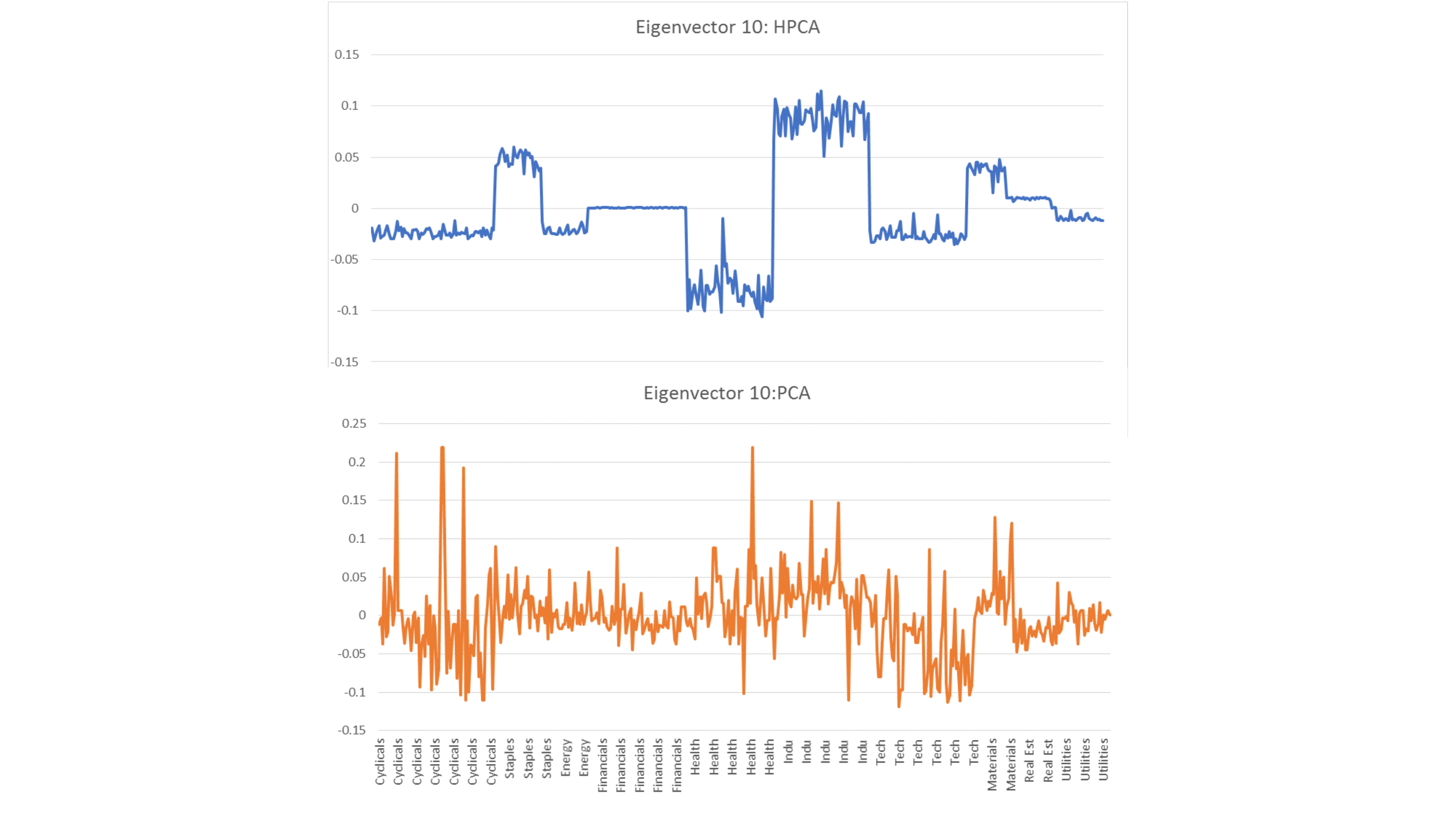} %, width=15 cm
\caption{Tenth eigenvectors.}
\label{ev10}
\end{bigcenter}
\end{figure}

For eigenvectors 2 through 5 Figures (\ref{ev2}) through (\ref{ev5}), we find that the PCA eigenvectors correspond to ``noisy versions'' of the corresponding HPCA eigenvectors. The latter are essentially long-short sector eigenportfolios. The discrepancy increases when we consider higher-order eigenvalues, beyond 5. Eigenvectors \#6 aren't similar as shown in Figure (\ref{ev6}). The PCA eigenvector contains both positive and negative signs within the Consumer Discretionary sector. Eigenvector 7 in HPCA is the first which is concentrated in a single sector, which is Consumer Discretionary (Fig. (\ref{ev7}). The remaining eigenvectors up to rank 10 are displayed in Figures (\ref{ev8}) to (\ref{ev10}).

The main conclusions are: (a) most of the top eigenvalues and corresponding eigenvectors are related
to the inter-sector correlation $\overline{\rho}$. This provides an interpretation for these eigenportfolios, or common risk factors, as ``portfolios of long-only sector portfolios''.
(b) The remaining eigenvectors may be quite different. The HPCA defines the factors into ``sector-sector'' and ``long-short intra-sector''. PCA eigenvectors, in contrast, become increasingly difficult to interpret as simple sector-sector interactions or intra-sector interactions.
\newpage
\newpage
\begin{figure}[htbp]
\begin{bigcenter}
\includegraphics[scale=0.30]{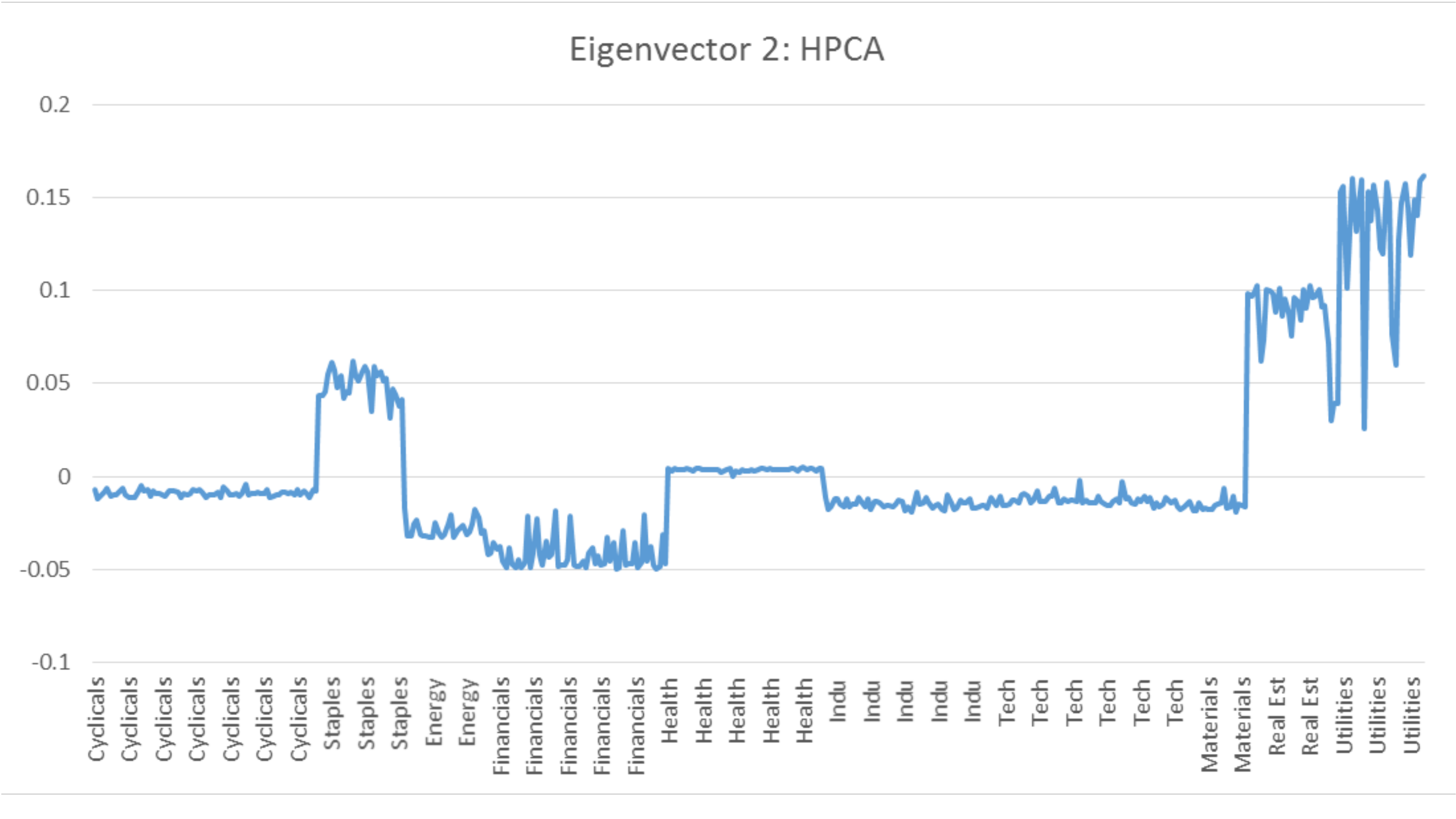} %, width=15 cm
\caption{Second eigenvector of HPCA. The variance explained is 4.7\% for HPCA and 6.1\% for PCA.}
\label{hpca2}
\end{bigcenter}
\end{figure}

\newpage

\section{Analysis of residuals via RMT \& Conclusion}
To further evaluate the HPCA, we considered both models (HPCA,PCA) with a cutoff $m=30$, and compared the multivariate statistics. We expect that after removing $m \approx 30$ eigenvectors, the correlations of the residuals (both intra- and inter- sector) should be small.

\begin{figure}[htbp]
\begin{bigcenter}
\includegraphics[scale=0.60]{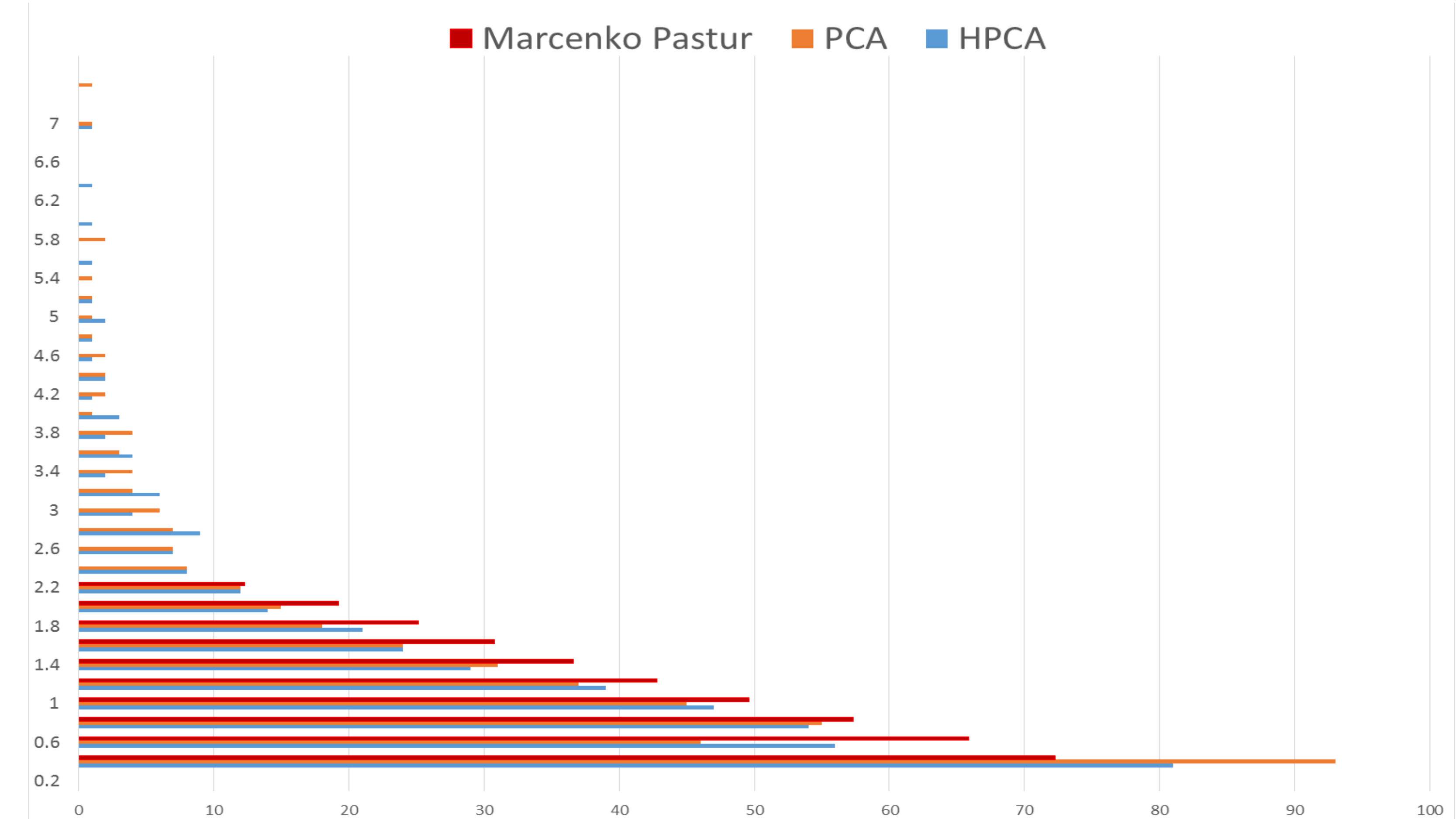} %, width=15 cm
\caption{Histograms of residuals for HPCA (blue) and PCA (orange) after removing the first 3 eigenportfolios. For reference we display the ``histogram'' of the Marcenko-Pastur (MP)
for corresponding to the same ratio of rows to columns ($1508 \times 434$). The histograms of HPCA and PCA are comparable. Both are localized below the critical MP level of 2.36,
with a smooth ``leakage'' as expected due to finite-size effects.}
\label{marcenko}
\end{bigcenter}
\end{figure}

Empirically, the top eigenvectors of the correlation matrices of residuals are approximately 6.8 (HPCA) and 7.7 (PCA), which correspond to an approximate average correlation of $7.3/434 =1.7\%$.  We compared the histograms of the eigenvalues for the corresponding correlation matrices and found that they are very near each other. We also compared the histograms with a discretization of the Marcenko-Pastur distribution (mimicking the comparable histogram for the large-matrix limit), suggesting that the residuals behave like a random matrix in both models; see Fig. (\ref{marcenko}). The majority of the lines, in both cases, are below the Marcenko-Pastur cutoff $\lambda^+=2.36$, as postulated by RMT, and have comparable sizes to the MP distribution. There are, nevertheless, some lines above the MP threshold in both models (which are essentially equal), but they decreasing in magnitude as $\lambda$ increases,and could perhaps be interpreted as finite-size fluctuations.

This calculation suggests that using the full empirical correlation matrix is not more informative than using the HPCA model, which uses only the sector correlation matrices, and in which intra-sector correlations are derived from the correlations of the EV1 for different sectors. Clearly, the HPCA provides a simpler description of common risk factors than PCA. The HPCA is therefore a viable alternative to PCA in the analysis of multivariate data in Finance, which should be of interest for asset-allocation and portfolio risk-management.

\end{document}